\documentclass[fleqn,twoside]{article}
\usepackage{espcrc2}

\usepackage{graphicx}
\usepackage[figuresright]{rotating}

\newcommand{\AmS}{{\protect\the\textfont2
  A\kern-.1667em\lower.5ex\hbox{M}\kern-.125emS}}

\title{Transiting topological sectors with the overlap}
\author{ 
Michael Creutz 
\address 
{Physics Department, Brookhaven
National Laboratory,
\\ Upton, NY 11973, USA
} 
\thanks
{This manuscript
has been authored under contract number DE-AC02-98CH10886 with the
U.S.~Department of Energy.  Accordingly, the U.S. Government retains a
non-exclusive, royalty-free license to publish or reproduce the
published form of this contribution, or allow others to do so, for
U.S.~Government purposes.  I benefited from discussions with Y. Shamir
and M. Golterman during visits made possible by grant No.~98-302 from
the United-States -- Israel Binational Science foundation (BSF),
Jerusalem, Israel.  }
}
       
\begin{document}

\begin{abstract}
The overlap operator provides an elegant definition for the winding
number of lattice gauge field configurations.  Only for a set of
configurations of measure zero is this procedure undefined.  Without
restrictions on the lattice fields, however, the space of gauge fields
is simply connected.  I present a simple low dimensional illustration
of how the eigenvalues of a truncated overlap operator flow as one
travels between different topological sectors.
\vspace{1pc}
\end{abstract}

\maketitle

\input epsf
\input colordvi  
\def \li {\par\hskip .1in \Green{$\bullet$}\hskip .1 in \textBlue}

\def \hi {\medskip \textBlack}
\def \topic #1{\par\textRed\centerline {#1}}

\long \def \blockcomment #1\endcomment{}


The overlap operator \cite{overlap} elegantly extends many features of
chiral symmetry to the lattice.  In particular, it provides a precise
definition of a ``winding number'' for gauge field configurations,
giving a lattice extention of continuum index theorems \cite{index}.
At first sight this seems remarkable since the space of Wilson gauge
fields is simply connected.  In selecting sectors, the overlap
operator must become singular at boundaries.  These singular
configurations form a set of measure zero.  A simple ``admissiblity''
criterion \cite{admissible} guarantees that the overlap operator is
well defined.  This criterion, however, is rather strong, and is not
generally satisfied for configurations in practical simulations.

Here I explore the behavior of the overlap operator as one passes
through a singularity separating two different sectors.  This requires
a truncation of the definition of the overlap.  The result is that two
complex eigenvalues of the overlap operator collide and evolve into a
zero mode plus one heavy real eigenvalue.  I follow this evolution
explicitly in a simple zero space-time dimensional toy model.

I briefly review the so called ``continuum'' Dirac action and the role
of zero modes.  The generic action for a gauge theory consists of a
pure gauge term and an interaction with the fermions, $S=S_g+S_f$.
The gauge part is the square of the field strength, $S_g={1\over 4}
\int F_{\mu\nu} F_{\mu\nu}$.  The fermion term is a quadratic form $
S_f=\int \overline\psi D_c \psi $ with
\begin{equation}
D_c=\gamma\cdot(\partial+ig A)+m
\end{equation}
The differential operator $D_c$ consists of an anti-Hermitean kinetic
term plus a Hermitean mass term.  It satisfies the Hermiticity condition
\begin{equation}
\gamma_5 D_c = D_c^\dagger\gamma_5
\end{equation}
Complex eigenvalues of $D_c$ are paired; if $D_c \chi = \lambda \chi$
then $D_c \gamma_5 \chi = \lambda^* \gamma_5 \chi$.  Restricting
ourselves to the space spanned by eigenvectors with real eigenvalues,
then $\gamma_5$ and $D_c$ can be simultaneously diagonalized.  On this
subspace, an integer index is the difference of the number of positive
and negative eigenvalues of $\gamma_5$, i.e. $\nu=n_+-n_-$.  This
number is robust under smooth field deformations and lies at the basis
of the index theorem, which says that this index can also be
calculated directly from the gauge fields as a topological charge
\cite{index}.

For comparison with the lattice theory, we can consider the free
continuum theory in momentum space $D_c= i p\cdot\gamma+m$.  This has
eigenvalues $\lambda=\pm i|p|+m$.  If we work in finite volume, the
momentum is quantized in units of ${2\pi\over L}$.  So called
``naive'' lattice fermions are obtained from the continuum result by
the simple substitution $p_\mu \rightarrow \sin(p_\mu a) / a$.  These
are fraught with the famous doublers, extra low energy states whenever
any component of the momentum satisfies $p_\mu \sim {\pi\over a}$.
The doubling problem was solved years ago by Wilson \cite{wilsonf},
who allowed the fermion mass to depend on momentum
\begin{equation}
m\rightarrow m+{1\over a} \sum_\mu (1-\cos(p_\mu a))
\end{equation}
thus giving the
doublers a mass of order $1/a$.  In momentum space, the free
Wilson-Dirac operator takes the form
\begin{equation}
D_w= m+{1\over a}\sum_\mu (i \sin(p_\mu a)\gamma_\mu+ 1-\cos(p_\mu a)).
\end{equation}

The difficulty with the Wilson approach is that the added term
violates chiral symmetry.  With gauge fields, the eigenvalues
drift. To maintain the physics of massless quarks requires fine
tuning.  Real eigenvalues of $D_w$ can appear along much of the real
axis, and for the purpose of defining an index we need a criterion for
which of them to include.

The overlap Dirac operator partially answers these questions
\cite{overlap}.  To construct this operator, one starts with $D_w$ at
a negative $m$.  This is projected onto a unitary matrix
\begin{equation}
V= D_w (D_w^\dagger D_w)^{-1/2}
\end{equation}
from this the overlap operator is simply
\begin{equation}
D=1+V
\end{equation}
Thus the low eigenvalues of $D_w$ become low eigenvalues of $D$, while
the higher ones are projected to the side of the unitarity circle near
unity.  This process is sketched here
\medskip
\epsfxsize \hsize
\centerline{\epsffile {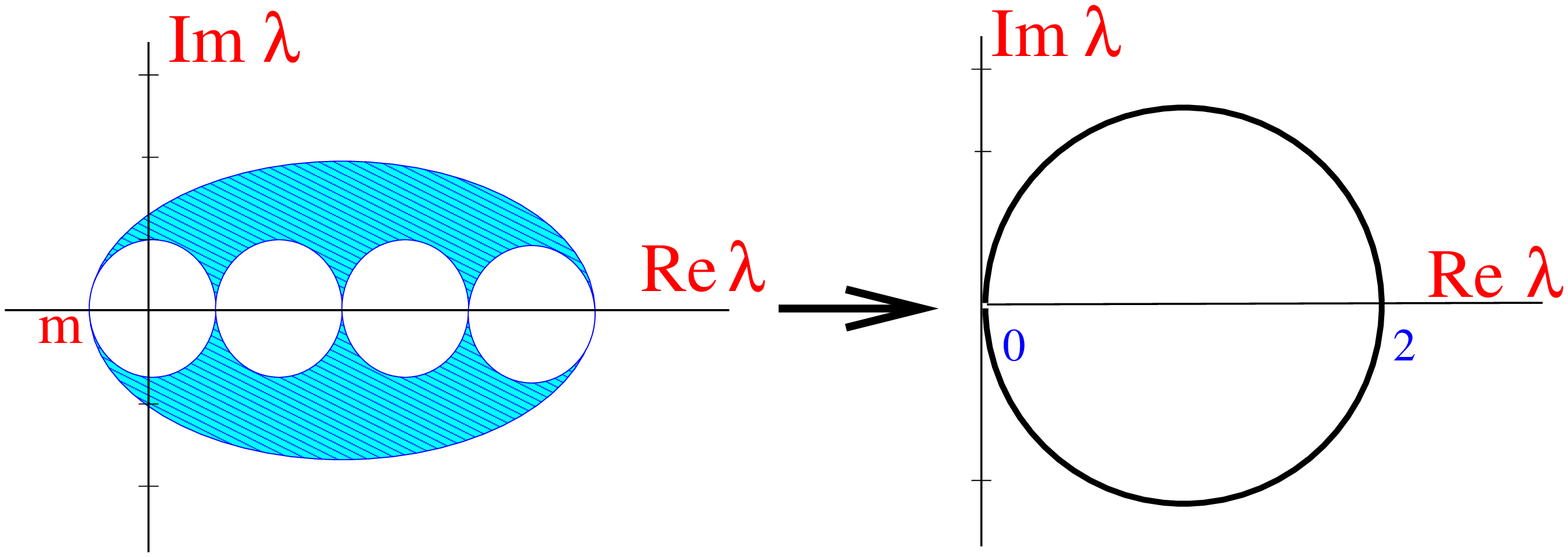}}

This construction satisfies the continuum property that $D$ is normal,
$\left[D,D^\dagger\right]=0$, and preserves the $\gamma_5$
Hermiticity, $\gamma_5 D = D^\dagger \gamma_5$.  Furthermore, we have
the famous Ginsparg-Wilson relation \cite{gw}, succinctly written as
\begin{equation}
D \gamma_5 = -\Gamma_5 D 
\end{equation}
with the new matrix
\begin{equation}
\Gamma_5 \equiv V\gamma_5 = (1-D)\gamma_5
\end{equation}
satisfying some of the same conditions as $\gamma_5$,
$\Gamma_5=\Gamma_5^\dagger$ and $\Gamma_5^2=\gamma_5^2=I$. The
eigenvalues of $\Gamma_5$ are all $\pm 1$, implying its trace is an
integer.  From this we define the gauge field index
\begin{equation}
\nu = {1\over 2} {\rm Tr\ }\Gamma_5
\end{equation}
Note the factor of 2, which comes from heavy modes at  $V \sim 1$.

Our fermionic action, $S_f=\overline \psi D \psi$, is invariant under
the generalized chiral rotation
\begin{equation}
\psi \rightarrow e^{i\theta\gamma_5}\psi \qquad
\overline\psi \rightarrow \overline\psi e^{i\theta\Gamma_5} 
\end{equation}
In this formalism the chiral anomaly appears in the fermionic measure
\begin{equation}
d\overline\psi\ d\psi \rightarrow 
d\overline\psi\ d\psi\ e^{-i{\rm Tr\ }\Gamma_5}
\rightarrow
d\overline\psi\ d\psi
\ e^{-2i\theta\nu}
\end{equation}
much as in continuum discussions \cite{fujikawa}.

A fermion mass introduces, as in the continuum, the possibility of a CP
violating term.  To have formulas similar to the continuum, it is
convenient to consider the fermionic action with mass term of the form
\begin{equation}
S_f=\overline \psi D \psi+ \overline\psi (1-V) M \psi/2
\end{equation}
The rotation $M \rightarrow e^{i\theta\gamma_5} M$ is physically
equivalent to a modification of the gauge action $S_g \rightarrow S_g
+ \theta\nu$.  This angle $\theta$ is the strong CP violating
parameter of innumerable continuum discussions.

This is all quite elegant, but the space of Wilson lattice gauge
fields is simply connected.  This raises the question of what happens
as one continues between topological sectors.  Along such a path $D$
must become singular.  To keep things well defined, I introduce a
cutoff into the definition of $V$
\begin{equation}
V= D_w (D_w^\dagger D_w+{\epsilon^2})^{-1/2}
\end{equation}
The quantity $\epsilon$ should be analogous to ${1\over L_5}$ with
domain wall fermions.

I now introduce a simple $2\times 2$ matrix example.  I take
effectively zero space-time dimensions, with $\sigma_3$ playing the
role of $\gamma_5$.  The hermiticity condition reduces to $
D_W^\dagger = \sigma_3 D_W \sigma_3.  $ The most general two by two
matrix satisfying this has the form
\begin{equation}
D_W=b_0+ib_1\sigma_1+ib_2\sigma_2 + b_3 \sigma_3
\end{equation}
This is singular when $|D_W|=b_0^2+b_1^2+b_2^2-b_3^2=0$.  We have a
Minkowski space with the role of time being played by $b_3$.
Minkowski space naturally breaks into light-like and space-like
sectors.  The index will highlight this division.

It is convenient to go to an analogue of ``polar'' coordinates and
reparametrize
\begin{equation}
D_W=U\ (a_0+a_3\sigma_3)\ U
\end{equation}
with $ U=e^{i(a_1\sigma_1+a_2 \sigma_2)/2}$.
The coordinate mapping is $b_3=a_3$ and
$a_0=\pm\sqrt{b_0^2+b_1^2+b_2^2}$.  We explicitly construct $V$
\begin{equation}\matrix{
V=D_W (D_W^\dagger D_W+\epsilon^2)^{-1/2}=\cr
 \cr
\ U\ \pmatrix{{a_0+a_3\over\sqrt{(a_0+a_3)^2+\epsilon^2}} & 0 \cr
         0         & {a_0-a_3 \over\sqrt{(a_0-a_3)^2+\epsilon^2}} \cr }
\ U \cr
}
\end{equation}

The possible topological sectors fall into three cases.  The first has
$a_0^2-a_3^2 > 0$ representing the spacelike sector of our Minkowski
space ($a_3=b_3$ plays the role of time).  In this case $V=U^2$ and
thus $D=1+U^2$.  This has a conjugate pair of eigenvalues $
\lambda_\pm= 1+e^{\pm i\sqrt{a_1^2+a_2^2}} $ The winding number
vanishes: $\nu={1\over 2}{\rm Tr\ } U\gamma_5 U^\dagger=0$

The second case involves $a_3 > |a_0|$.  Then $V=\sigma_3 $ and $D=
1+\sigma_3$.  The winding number $\nu=1$; so, this represents the
analog of an {``instanton''}.  The third and final case is a
reflection of this, with $a_3 < -|a_0|$, $D= 1-\sigma_3$, and
$\nu=-1$.

Now I transit between these sectors.  As an example, let $a_3$ pass
through the ``light cone'' at $a_0>0$.  To be explicit, use
$U=c-is\sigma_2$ with $c^2+s^2=1$.  For our interpolation parameter,
define $x={a_0-a_3 / \sqrt {(a_0-a_3)^2+\epsilon^2}}$ with range
$-1\le x \le 1$.  With the cutoff in place $V$ is no longer unitary,
but takes the form
\begin{equation}
V={1\over 2}\pmatrix{1+c-x+cx & -s-sx \cr
            s+sx & -1+c+x+cx \cr}  
\end{equation}
The eigenvalues of this are
\begin{equation}
\lambda={1\over 2}\left ( c(1+x) \pm \sqrt{c^2(1+x)^2-4x}\right )
\end{equation}
As an eigenvalue of $D_W^\dagger D_W$ passes through zero, a pair of
eigenvalues leaves the unitarity circle.  This is a perpendicular
departure, following another circle.  These eigenvalues then collide
and become real.  They move out to rest at $\pm 1$.  In the process
the winding number changes by one unit.  This behavior is sketched
here
\bigskip
\epsfxsize .45 \hsize
\centerline{\epsffile {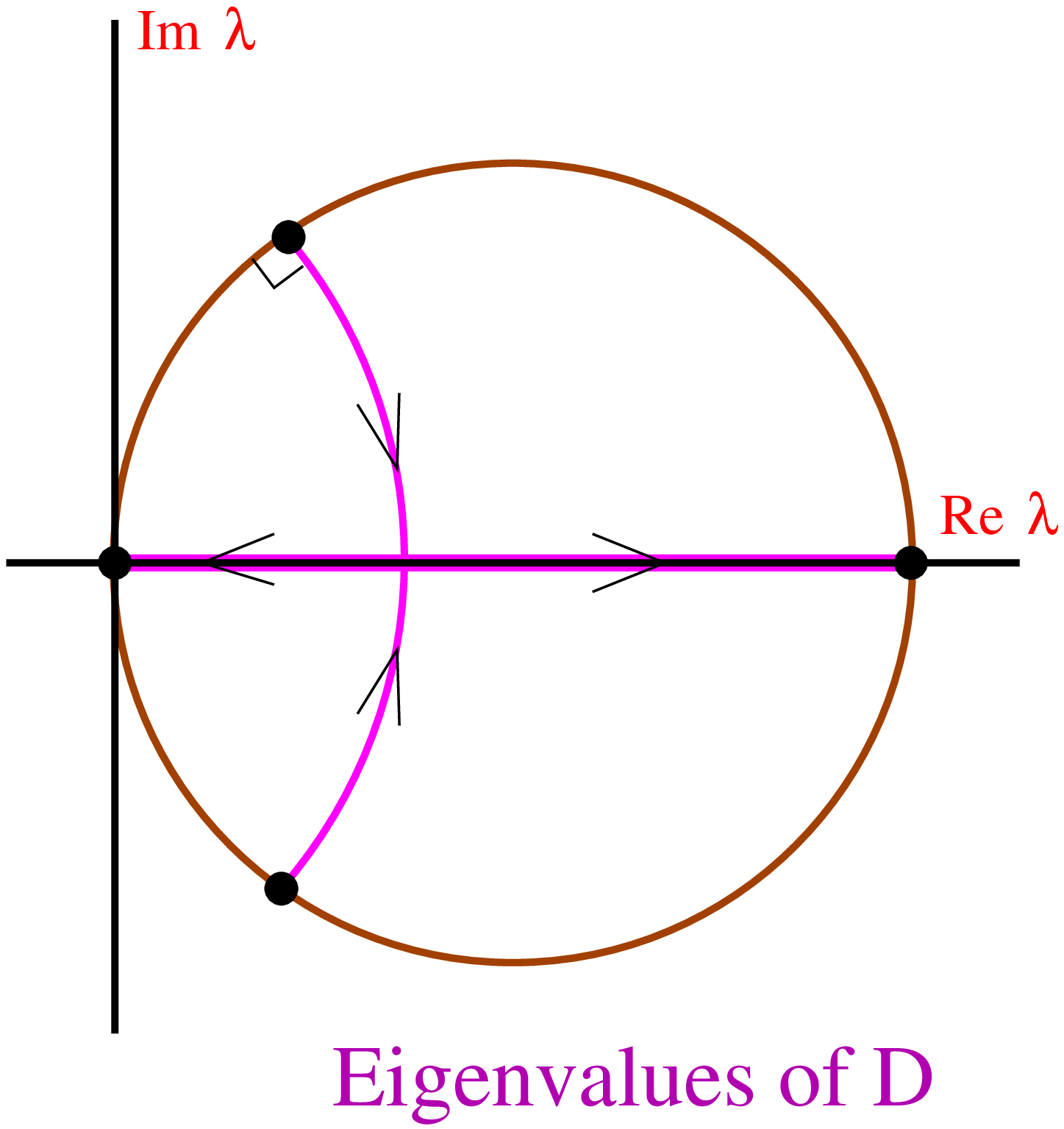}}

The participating eigenvalues can come from anywhere on the unitarity
circle.  An instanton falling through the lattice does not require a
large fermionic action.  Throughout this discussion the index $\nu$ is
an integer except within $\epsilon$ of sector boundaries.  This
behavior is fairly robust, with other eigenvalues of $V$ moving
little.  However, as shown in \cite{lang} the other eigenvalues can
also briefly leave the unitarity circle as we pass through the
boundary.

\end{document}